\documentclass[a4paper,11pt,oneside]{article}

\usepackage[english]{babel}
\usepackage[T1]{fontenc}
\usepackage[utf8]{inputenc}

\usepackage[pdftex,unicode]{hyperref}
\hypersetup{pdftitle=Are Economically Advanced Countries More Efficient in Basic and Applied Research?}
\hypersetup{pdfauthor=Vladimír Holý and Karel Šafr}

\usepackage{xcolor}
\definecolor{mycolor}{rgb}{0.04,0.20,0.40}
\hypersetup{colorlinks=true, linkcolor=mycolor, anchorcolor=mycolor, citecolor=mycolor, filecolor=mycolor, urlcolor=mycolor}

\usepackage[margin=60pt]{geometry}
\usepackage{amsmath}
\usepackage{amssymb}
\usepackage{bm}
\usepackage{amsthm}
\newtheorem{hypothesis}{Hypothesis}

\usepackage{enumitem}

\usepackage{graphicx}
\usepackage{float}
\usepackage{hhline}
\usepackage{multirow}

\usepackage[authoryear]{natbib}

\begin{document}

\begin{center}
{\Large \bfseries Are Economically Advanced Countries More Efficient in Basic and Applied Research?}\footnote{Preliminary results of data envelopment analysis not accounting for purchasing power parities were presented in \cite{Holy2017k}.}
\end{center}

\begin{center}
{\bfseries Vladimír Holý} \\
University of Economics, Prague \\
Winston Churchill Square 4, 130 67 Prague 3, Czech Republic \\
\href{mailto:vladimir.holy@vse.cz}{vladimir.holy@vse.cz} \\
Corresponding Author
\end{center}

\begin{center}
{\bfseries Karel Šafr} \\
University of Economics, Prague \\
Winston Churchill Square 4, 130 67 Prague 3, Czech Republic \\
\href{mailto:karel.safr@vse.cz}{karel.safr@vse.cz}
\end{center}

\begin{center}
{\itshape April 25, 2018}
\end{center}

\noindent
\textbf{Abstract:}
Research and development (R\&D) of countries play a major role in a long-term development of the economy. We measure the R\&D efficiency of all 28 member countries of the European Union in the years 2008--2014. Super-efficient data envelopment analysis (DEA) based on robustness of classification into efficient and inefficient units is adopted. We use the number of citations as output of basic research, the number of patents as output of applied research and R\&D expenditures with manpower as inputs. To meet DEA assumptions and to capture R\&D characteristics, we analyze a homogeneous sample of countries, adjust prices using purchasing power parity and consider time lag between inputs and outputs. We find that the efficiency of general R\&D is higher for countries with higher GDP per capita. This relation also holds for specialized efficiencies of basic and applied research. However, it is much stronger for applied research suggesting its outputs are more easily distinguished and captured. Our findings are important in the evaluation of research and policy making.
\\

\\

\\

\section{Introduction}
\label{sec:intro}

\emph{Research and development (R\&D)} is work directed towards corporate or government innovation. We focus on R\&D on the national level as it plays a major role in the development of the economy in long term. For that reason, many countries invest significant part of their resources to R\&D. These expenditures are directly a part of the gross domestic product and immediately contribute to economic growth. However, the purpose of R\&D is its long-term effect on economic growth. In general, R\&D can be classified into \emph{basic research} aiming at improvement of scientific theories and \emph{applied research} aiming at developing new technologies.

The question is whether R\&D operations in a country are efficient in comparison with other countries. We measure R\&D efficiency by \emph{data envelopment analysis (DEA)}. Our main goal is to determine the efficiencies of basic and applied research within European Union countries and to examine the following two hypotheses:

\begin{hypothesis}
The efficiency of a country's R\&D is interlinked with its level of economic advancement. The greater the economic advancement of a country the more efficient its R\&D.
\end{hypothesis}

\begin{hypothesis}
The relation between R\&D efficiency and the level of economic advancement of a country also holds separately for basic and applied research. The relation is stronger for applied research as its outputs are easier to capture and more conspicuous.
\end{hypothesis}

We review the literature related to R\&D in Section \ref{sec:litRad}. We focus mainly on differences between basic and applied research as well as on possible proxy variables. Advancements of DEA are highlighted in Section \ref{sec:litDea}. There are many studies analyzing the R\&D efficiency of countries using DEA methodology. We survey the literature regarding R\&D efficiency in Section \ref{sec:litEff}. However, many of these studies overlook the importance of analyzing the efficiency within the context of basic and applied research.

To rank countries by their R\&D efficiency we adopt \emph{Chebyshev distance DEA} proposed by \cite{Hladik2019}. This method has many advantages over the classical DEA. It is based on robustness of the (in)efficiency, is invariant to scaling, measures super-efficiency and is naturally normalized. We describe this DEA model in Section \ref{sec:methDea}. To examine our hypotheses we use panel regression with random time effects. In this model, the limited value of the efficiency is captured by logistic function. We describe this panel regression model in Section \ref{sec:methPanel}.

In the empirical study, we compare R\&D efficiency of 28 countries of European Union in the years 2008--2014. As inputs to the R\&D of a country we have chosen \emph{total R\&D expenditures (TERD)} and the \emph{number of scientist and engineers (SAE)}. The output of basic research is proxied by the \emph{number of citations of scientific publications (CIT)} while the output of applied research is proxied by the \emph{number of patent applications granted to the European Patent Office (EPO)}. The level of economic advancement of a country is represented by the \emph{gross domestic product per capita (GDP)}. We discuss the availability of the data and selection of variables in Section \ref{sec:empData}. Other possible variables are discussed in Section \ref{sec:empOther}. In Section \ref{sec:empDea}, we rank countries based on overall R\&D efficiency as well as specialized efficiencies for basic and applied research. We compare our results of efficiency analysis with other studies in Section \ref{sec:empComp}. The relation of R\&D efficiency and the level of economic advancement of a country is investigated in Section \ref{sec:empPanel}. We find that the overall efficiency is higher for countries with higher GDP supporting Hypothesis 1. We also find that the relation of applied research efficiency to GDP is more distinctive than the one of basic research efficiency supporting Hypothesis 2. Implications of our findings are discussed in Section \ref{sec:empImpl}.

We conclude the paper in Section \ref{sec:con}. The first contribution of our paper is the ranking of European countries based on basic and applied research that takes into account homogeneity of countries, purchasing power parity and the time delay between inputs and outputs. The second contribution is the exposure of positive relation between the R\&D efficiency and the GDP of a country. The third and final contribution is the assessment of differences between basic and applied research in the context of R\&D efficiency.

\section{Literature Review}
\label{sec:lit}

\subsection{Basic vs. Applied Research}
\label{sec:litRad}

\cite{Berbegal-Mirabent2015} studied interactions between basic and applied research at Spanish universities in 2006-2010 using a regression model. They used the number of scientific publications as a proxy for basic research and the number of patents as a proxy for applied research. A study conducted by \cite{Oecd2004} about the effects of patents showed that universities started to focus more on applied research when the number of patents was included as a factor in budget allocation. \cite{Schmoch2004} studied institutional research in the European Union. Again, he used the number of scientific publications as proxy for basic research and the number of patents as a proxy for applied research. However, he discussed that these variables might not be ideal as not all universities are registered as patent applicants. \cite{Guellec2005} noted that the use of patents as an indicator in R\&D refers to applied research rather than to basic research.

The classification into basic and applied research was also criticized. \cite{Narayanamurti2013} argued that it does not reflect rich connections between various types of research. \cite{Desrochers1998} warned that the patents are hard to compare as their real value can vary. They also do not have to be the outcome of R\&D but a defensive strategy of a company. Finally, a lot of R\&D outcome is impossible to patent or is simply not patented due to secrecy or morality reasons. \cite{Krejci2016} dealt with a scientific monographs evaluation with a focus on the subjectivity in the peer-review process. Problems with bibliometric data may arise for example due to a language bias as argued by \cite{Leeuwen2001}.

\subsection{Data Envelopment Analysis}
\label{sec:litDea}

DEA is a non-parametric method measuring efficiency of \emph{decision making units (DMU)}. It was introduced by \cite{Charnes1978}. In this model (also known as the CCR model), \emph{constant returns to scale} are assumed. \cite{Banker1984} proposed a model with \emph{variable returns to scale} (also known as the BCC model). \cite{Dyson2001} discuss a range of pitfalls and protocols in DEA applications such as homogeneity assumptions, returns to scale assumptions, input and output variables selection and measurement problems. Since the original DEA study there has been published a considerable amount of papers about DEA theory and applications. \cite{Emrouznejad2008} list the DEA publications in the first 30 years of its existence.

Classical DEA ranks only inefficient DMUs, the efficient DMUs have all the same efficiency score equal to one. This shortcoming is overcome in many modifications of DEA model. The ranking of efficient units is known as \emph{super-efficiency}. This concept was introduced by \cite{Andersen1993}. \cite{Banker2017} discuss the use of super-efficiency in identification of outliers. Other DEA models of super-efficiency were proposed for example by \cite{Tone2002}, \cite{Jablonsky2012} and \cite{Hladik2019}.

Another important issue is the sensitivity and stability of DMUs classification \citep{Charnes1992, Jahanshahloo2011}. \cite{Cooper2001} surveyed analytical methods that study variations of single DMU as well as all DMUs. \cite{Hladik2019} took a different approach and proposed DEA model based directly on robustness of (in)efficiency.

There are many methods for the subsequent analysis of the dependence between the efficiency score and relevant variables. This investigation of the exogenous influence is known as the second stage of DEA analysis. The most common approach is to use the logistic transformation in a regression model \citep{Papke1996}. Other approaches include the beta regression and the fractional regression \citep{Ramalho2010}. The two-stage double bootstrap DEA is also based on a regression model and allows to evaluate potential bias in the efficiency score \citep{Simar2007, Halkos2013}. The non-parametric conditional DEA can be used to measure trade-off between efficiency and other variables \citep{Varabyova2017}.

\subsection{Efficiency of Research and Development}
\label{sec:litEff}

Analysis of R\&D efficiency using DEA is the subject of many studies. \cite{Nasierowski2003} used DEA to measure the influence of R\&D efficiency on productivity of a country by Tobit regression. \cite{Lee2005} calculated R\&D efficiency as well as specialized efficiencies for 27 countries around the world grouped into 4 clusters to achieve homogeneity. \cite{Wang2007a} analyzed efficiency of aggregated R\&D activities of 30 countries using the R\&D capital stock and the manpower as inputs and the number of patents and the number of scientific publications as outputs. \cite{Sharma2008} studied efficiency of 22 countries using R\&D expenditures and number of scientists as inputs and the number of patents as output. \cite{Lee2009} compared national R\&D programs using DEA. \cite{Roman2010} analyzed regional efficiencies of Romania and Bulgaria. \cite{Thomas2011} studied patent efficiencies of U.S. states with a three-year delay between inputs and outputs. \cite{Aristovnik2012} studied the efficiency of R\&D as well as education focusing on the new EU member states. \cite{Cullmann2012} used a two-stage semi-parametric DEA to analyze R\&D. \cite{Han2016} analyzed efficiency of R\&D of 15 Korean regions. \cite{Jablonsky2016} evaluated both research and teaching performance of economic faculties in the Czech Republic.

\cite{Cincera2009} analyzed R\&D efficiencies of OECD countries using DEA and stochastic frontier analysis in an extensive study. To rank countries they utilized financial indicators with variable returns to scale. However, they report that the results vary for DEA and stochastic frontier analysis because of the different specifications of models.

\cite{Chen2011} used DEA to rank 24 countries around the world. They used R\&D manpower and expenditures as inputs and the number of patents, the number of scientific articles and royalty and licensing fees as outputs. They find that the R\&D intensity, intellectual property rights protection, knowledge stock and human capital accumulation have positive effects on R\&D efficiency.

\cite{Ekinci2015} surveyed the literature concerning with R\&D efficiency and discussed the selection of input and output variables for DEA model. \cite{Ekinci2017} followed with a study of R\&D efficiency of European Union countries. They use 9 variables in total. As inputs they used R\&D expenditures conducted by business enterprises, government and higher education sector, the number of full time R\&D personnel, the number of people with tertiary education employed in science and technology and employment in high technology and knowledge-intensive sectors. As outputs they used the number of scientific publications, the number of patents granted by the European Patent Office (EPO) and the number of patents granted by the United States Patent and Trademark Office.

\section{Methodology}
\label{sec:meth}

\subsection{Chebyshev Distance DEA}
\label{sec:methDea}

Let us assume we have $I$ input variables and $J$ output variables of $C$ countries in $T$ times. In general, DEA measures efficiency of a given DMU relative to other DMUs in the set based on how efficiently it can transform inputs into outputs. In our case, we consider different times $t$ as different sets (i.e. we compare a given DMU to other DMUs in the same time). For every DMU given by a country $c$ and a time $t$, the problem of measuring efficiency is formulated as finding the optimal weights of input and output variables with respect to other DMUs at time $t$. We also consider the possibility of a time delay between inputs and outputs. In many real-word situations, it takes some time for inputs to be transformed into outputs. For example a scientific paper is published in a journal (output) after one or two years its authors received the funding (input). For that reason we take outputs from time $t$ but inputs are taken from time $t-h$ where $h \geq 0$ is the time delay.

To evaluate efficiency of DMUs we adopt the Chebyshev distance DEA with constant returns to scale proposed by \cite{Hladik2019}. The idea behind this method is to rank DMUs based on robustness of efficiency or inefficiency classification to variations of input and output data using Chebyshev distance. Let $\bm{X}_t = (x_{c,t,i})_{c=1,i=1}^{C,I}$ be the nonnegative matrix of inputs and $\bm{Y}_t = (y_{c,t,j})_{c=1,j=1}^{C,J}$ be the nonnegative matrix of outputs in a time $t$. We denote $\bm{x}_{c,t}$ and $\bm{y}_{c,t}$ the vectors corresponding to the $c$-th row. We also denote $\bm{X}_{-c,t}$ and $\bm{Y}_{-c,t}$ the matrices with $c$-th row missing (i.e. the inputs and outputs of every DMU but $c$). For DMU given by country $c=1,\ldots,C$ and time $t=h+1,\ldots,T$, the efficiency score $r_{c,t} = 1 + 2 \delta_{c,t}^*$ is given by the optimal solution $\delta_{c,t}^*$ to the optimization problem
\begin{equation}
\begin{aligned}
\max_{\substack{\delta_{c,t}, \\ \bm{\mu}_{c,t}, \bm{\nu_{c,t}}}} && \multispan2{$\delta_{c,t}$} \\
\text{s. t.} && (1 - \delta_{c,t}) \bm{y}_{c,t}' \bm{\mu}_{c,t} & \geq 1, \\
             && (1 + \delta_{c,t}) \bm{x}_{c,t-h}' \bm{\nu}_{c,t} & \leq 1, \\
             && (1 + \delta_{c,t}) \bm{Y}_{-c,t} \bm{\mu}_{c,t} - (1 - \delta_{c,t}) \bm{X}_{-c,t-h} \bm{\nu}_{c,t} & \leq \bm{0}, \\
             && \bm{\mu}_{c,t} & \geq \bm{0}, \\
             && \bm{\nu}_{c,t} & \geq \bm{0}, \\
\end{aligned}
\end{equation}
where $\bm{\nu_{c,t}} = (\nu_{c,t,1}, \ldots, \nu_{c,t,I})'$ are weights of inputs and $\bm{\mu_{c,t}} = (\nu_{c,t,1}, \ldots, \nu_{c,t,J})'$ are weights of outputs. We use (perhaps unnecessary) notation of indices $c$ and $t$ to emphasize that every country in every time has its own weights. The above introduced model is a nonlinear programming problem. However, it can be effectively approximated by linear programming problem of the form
\begin{equation}
\begin{aligned}
\max_{\substack{\delta_{c,t}, \\ \bm{\mu}_{c,t}, \bm{\nu_{c,t}}}} && \multispan2{$\delta_{c,t}$} \\
\text{s. t.} && \bm{y}_{c,t}' \bm{\mu}_{c,t} & \geq 1 + 2 \delta_{c,t}, \\
             && \bm{x}_{c,t-h}' \bm{\nu}_{c,t} & \leq 1 - 2 \delta_{c,t}, \\
             && \bm{Y}_{-c,t} \bm{\mu}_{c,t} - \bm{X}_{-c,t-h} \bm{\nu}_{c,t} & \leq \bm{0}, \\
             && \bm{\mu}_{c,t} & \geq \bm{0}, \\
             && \bm{\nu}_{c,t} & \geq \bm{0}. \\
\end{aligned}
\end{equation}
Values $r_{c,t} \in (0,1)$ indicate inefficient DMU while values $r_{c,t} \in [1,2)$ indicate efficient DMU. 

The Chebyshev distance DEA has many desirable properties. As in classical DEA, the efficiency scores are invariant to scaling of variables. The classification into efficient and inefficient DMUs as well as the order of inefficient DMUs according to their efficiency score is exactly the same as in classical DEA. The difference between methods is in the values of efficiency scores. However, unlike the classical DEA, the Chebyshev distance DEA also ranks efficient DMUs. This is related to the super-efficiency of \cite{Andersen1993} with a key difference that their efficiency scores for efficient DMUs are not invariant to scaling. The efficiency scores given by the Chebyshev distance DEA are also naturally normalized which allows the comparison of DMUs between different models.

Overall, the Chebyshev distance DEA is similar to the classical DEA model of \cite{Charnes1978} with the super-efficiency extension (achieved by excluding DMU under evaluation from the creation of efficiency frontier) of \cite{Andersen1993}. The Chebyshev distance DEA offers quite straightforward interpretation. For an inefficient DMU, the efficiency score is the Chebyshev distance to the nearest efficient point (smallest possible variation of all inputs and outputs causing efficiency), while for an efficient DMU, the efficiency score is the Chebyshev distance to the nearest inefficient point (largest possible variation of all inputs and outputs preserving efficiency). The efficiency score directly indicates how a DMU is sensitive to any changes of its inputs and outputs. Therefore, the sensitivity analysis of \cite{Charnes1992} based on regions of stability is contained in efficient scores. A region of stability is a cell in which all perturbations of inputs and outputs based on a given measure (the Chebyshev distance in our case) preserve the efficiency/inefficiency of considered DMU while the other DMUs are unchanged. Stable inefficient DMUs have low efficiency score, stable efficient DMUs have high efficiency score and unstable DMUs have efficiency score around 1.

\subsection{Panel Regression with Random Time Effects and Logistic Function}
\label{sec:methPanel}

Our goal is to find a relation between the efficiency score $r_{c,t}$ of a country $c=1,\ldots,C$ in a time $t=1,\ldots,T$ and some other $K$ variables denoted by vector $\bm{z}_{c,t}=(z_{c,t,1},\ldots,z_{c,t,K})'$. Generally, the efficiency score has limited values $r_{c,t} \in (0,s)$. In our case of the Chebyshev distance DEA, $r_{c,t} \in (0,2)$, i.e. $s=2$. To capture limited values of efficiency score we use logistic function. The logistic function is rotationally symmetric around its inflection point (in our case 1). This makes it suitable for modeling of efficiency score by the Chebyshev distance DEA as it is also symmetric around 1 in the sense that it is based on robustness to (in)efficiency (i.e. the distance to 1 for inefficient DMUs as well as efficient DMUs given by variations of data).

As we have efficiency scores for $C$ countries and $T$ times we utilize panel regression with random time effects \citep{Baltagi2013}. We combine panel regression with logistic function into model
\begin{equation}
r_{c,t} = \frac{s}{1 + \exp (\alpha - \bm{z}_{c,t}' \bm{\beta} + u_t + v_{c,t}) }, \quad  \text{for } c = 1,\ldots,C, \ t = 1, \ldots, T,
\end{equation}
where $u_t \stackrel{iid}{\sim} \mathrm{N} (0, \sigma^2)$ are random time effects, $v_{c,t} \stackrel{iid}{\sim} \mathrm{N} (0, \omega^2)$ are idiosyncratic error terms, $\alpha$ is the intercept and $\bm{\beta} = (\beta_1,\ldots,\beta_K)'$ is the vector of coefficients. Notice that $\alpha$ as well as $\bm{\beta}$ are the same for every country and every time. In this model, we capture fluctuations of GDP level in time by the random time effects $u_t$. The above model can be linearized to
\begin{equation}
\ln \left( \frac{s}{r_{c,t}} - 1 \right) = \alpha - \bm{z}_{c,t}' \bm{\beta} + u_t + v_{c,t}, \quad  \text{for } c = 1,\ldots,C, \ t = 1, \ldots, T,
\end{equation}
which is a standard model of panel regression with random time effects (only with transformed dependent variable).

\section{Empirical Study}
\label{sec:emp}

\subsection{Data and Variable Selection}
\label{sec:empData}

In the empirical study, we analyze member countries of the European Union in the years 2007--2014. The reason for including only European Union countries in the study is their homogeneity in the sense of a common law and economic environment and single aim of R\&D. Although, some authors ignore homogeneity of countries \citep{Cincera2009, Chen2011}, it is an important assumption of DEA as stressed for example by \cite{Dyson2001} for general DEA applications and \cite{Lee2005} specifically for national R\&D efficiency. We use the following variables:

\begin{description}
\item[\textbf{Total R\&D expenditures (TERD)}] As one of the inputs, we use the total intramural expenditures to R\&D. Intramural expenditures include expenditures regardless of the source of funds in the country and also transactions outside the country. We use prices in Euros adjusted to constant prices of the year 2010 to ensure consistency of the data in time. Next, we adjust the prices using purchasing power parities to ensure comparability between countries with regard to different purchasing power. Most of R\&D efficiency studies ignore this step and we expect it to have a major impact on efficiency of countries with lower R\&D expenditures and lower prices.
\item[\textbf{Number of scientist and engineers (SAE)}] Another input is the number of scientist and engineers. This variable includes persons with scientific or technological training who are engaged in professional work on science and technology activities, high-level administrators and personnel who direct the execution of science and technology activities.
\item[\textbf{Number of citations of publications (CIT)}] As output variable of basic research, we use the number of citations of scientific publications. More specific, it is the number of citations recieved to all scientific documents until the given year.
\item[\textbf{Number of patents (EPO)}] With regard to inclusion of countries only from the European Union we use the number of patent applications granted to the European Patent Office in a given year as output variable of applied research.
\item[\textbf{Gross domestic product per capita (GDP)}] As proxy variable for the level of economic advancement of a country, we use gross domestic product per capita in constant prices of the year 2010 adjusted by purchasing power parities. This is the same approach as with TERD variable.
\end{description}

The data source of TERD, SAE, EPO and GDP variables is Eurostat. The data source of CIT variable is Scimago Journal \& Country Rank which aggregate data from Scopus.

\subsection{Discussion About Other Variables}
\label{sec:empOther}

We have also considered other variables as inputs and outputs. Besides SAE we have considered the number of people with tertiary education employed in science and technology. However, this variable did not have any notable impact on results and we have decided to use only SAE variable. We have also considered classification of R\&D expenditures to government expenditures and business expenditures. However, we have decided for total expenditures to keep the number of variables in DEA model small. As inputs we have also considered h-index and high-tech share on export. However, these variables are indices and therefore not very suitable for DEA analysis when mixing with nominal values \citep{Dyson2001}.

\subsection{Results of Efficiency Analysis}
\label{sec:empDea}

We apply the Chebyshev distance DEA separately for every year from 2008 to 2014. As it takes some time for inputs to be transformed into outputs, we use a one-year delay (i.e. inputs are taken from the previous year than outputs). We also assume constant returns to scale.

Table \ref{tab:effGeneral} reports efficiencies of general R\&D for 28 European Union countries in the years 2008--2014. With the exception of Latvia the results do not significantly change from year to year. This indicates that efficiencies are stable in time and without any major trends (at least in the studied period). As we evaluate efficiencies in each year separately we cannot observe the effects of crisis or other effects connected to economic cycles or supra-national R\&D programs. The use of data adjusted to purchasing power parities proved to be very significant in our analysis. This was evident specially for countries with lower R\&D expenditures and lower prices which have lower and more reasonable efficiencies than in the case of original data. Similar results are reported for efficiencies of basic research in Table \ref{tab:effBasic} and applied research in Table \ref{tab:effApplied}.

The efficiency score can also be interpreted as a sensitivity to changes of inputs and outputs. We can see in Table \ref{tab:effGeneral} that Denmark, Greece, Italy, Netherlands and Sweden have efficiency score close to 1 and therefore are very sensitive to changes in variables in the year 2014. A small variation in their inputs or outputs can lead to reclassification from efficient to inefficient and vice versa. Contrary, Cyprus and Finland are stable in their efficiency while the remaining countries are stable in their inefficiency in the year 2014.

\begin{figure}
\begin{center}
\includegraphics[width=118mm]{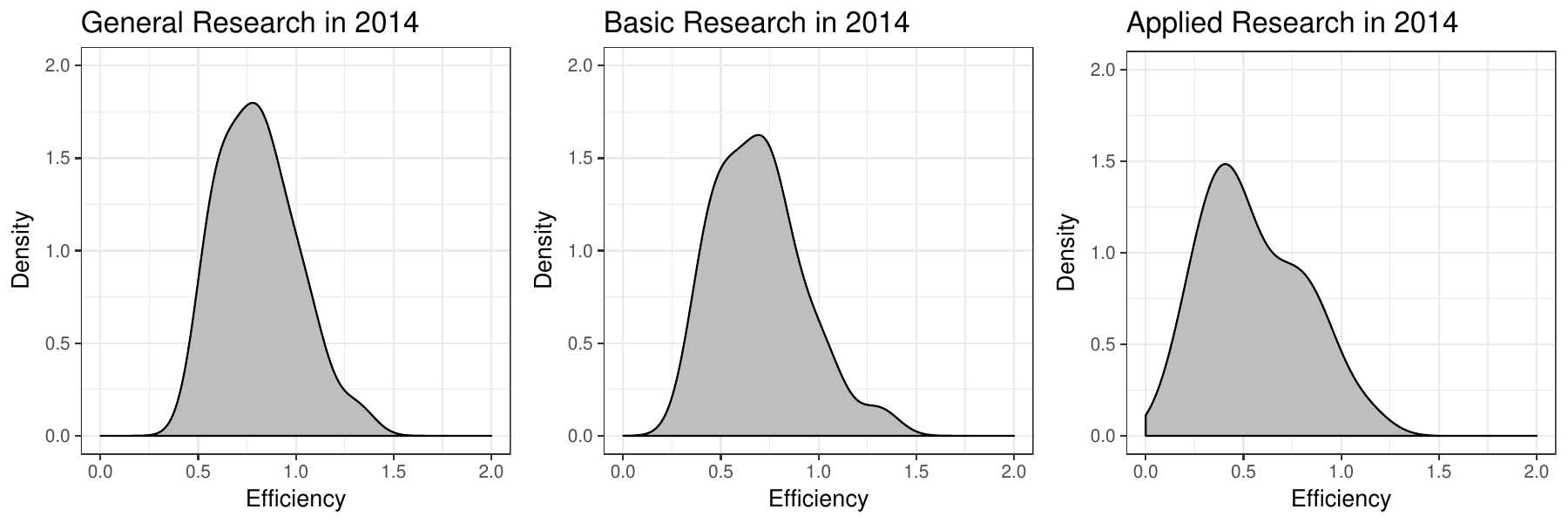} 
\caption{Kernel densities of research efficiencies of European Union countries in 2014.}
\label{fig:density}
\end{center}
\end{figure}

\begin{figure}
\begin{center}
\includegraphics[width=118mm]{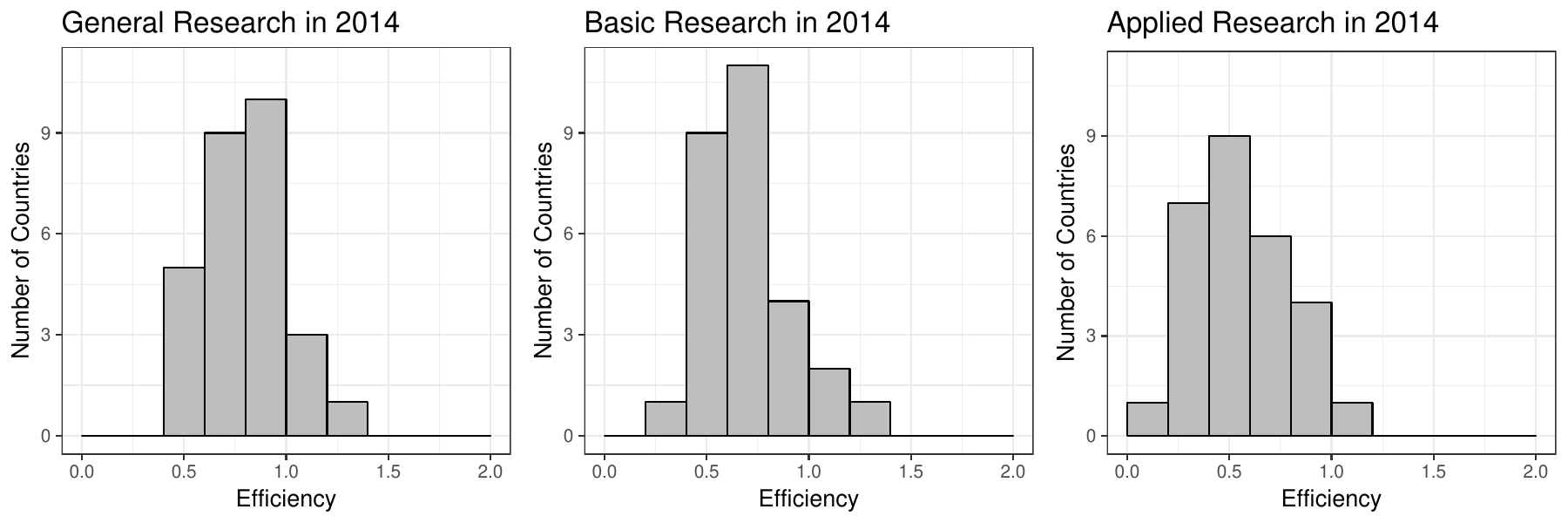} 
\caption{Histograms of research efficiencies of European Union countries in 2014.}
\label{fig:hist}
\end{center}
\end{figure}

Figure \ref{fig:density} contains kernel densities while Figure \ref{fig:hist} contains histograms of efficiencies and specialized efficiencies in the year 2014. We can see the differences between basic and applied research as 8 countries have the efficiency lower than 0.4 in applied research and only Lithuania has the efficiency lower than 0.4 in basic research. This means that the differences between countries in basic research are not as significant as in applied research. 

We take a closer look at the case of Greece. As we can see in Table \ref{tab:effBasic} and Table \ref{tab:effApplied}, Greece is efficient in basic research while it performs very poorly in applied research. Overall, it is efficient according to Table \ref{tab:effGeneral}. Greece has comparable bibliometrics to the European average per capita. After the economic crisis in the 2009 there was a decline in GDP and a significant drop in R\&D expenditures. This was followed by a decline in the number of published articles. As both R\&D expenditures and the number of published articles decreased, the efficiency score remained more or less on the same level. This suggests a good choice of 1 year as the time delay.

\begin{table}
\begin{center}
\begin{tabular}{llrrrrrrr}
\hline
Code & Country & 2008 & 2009 & 2010 & 2011 & 2012 & 2013 & 2014 \\
\hline
AT & Austria & 0.971 & 0.987 & 0.928 & 0.881 & 0.919 & 0.849 & 0.805 \\ 
BE & Belgium & 0.891 & 0.851 & 0.860 & 0.812 & 0.884 & 0.870 & 0.826 \\ 
BG & Bulgaria & 0.764 & 0.636 & 0.586 & 0.558 & 0.632 & 0.533 & 0.586 \\ 
HR & Croatia & 0.651 & 0.758 & 0.801 & 0.894 & 0.930 & 0.772 & 0.882 \\ 
CY & Cyprus & 1.039 & 1.156 & 1.148 & 1.129 & 1.258 & 1.259 & 1.310 \\ 
CZ & Czech Republic & 0.574 & 0.608 & 0.633 & 0.639 & 0.599 & 0.574 & 0.552 \\ 
DK & Denmark & 1.071 & 1.029 & 1.173 & 1.172 & 1.103 & 1.032 & 1.060 \\ 
EE & Estonia & 0.833 & 0.763 & 0.854 & 0.736 & 0.605 & 0.612 & 0.638 \\ 
FI & Finland & 0.854 & 0.879 & 0.935 & 0.873 & 1.046 & 1.129 & 1.133 \\ 
FR & France & 0.820 & 0.838 & 0.812 & 0.821 & 0.840 & 0.825 & 0.771 \\ 
DE & Germany & 1.108 & 1.133 & 1.106 & 1.041 & 0.995 & 0.954 & 0.886 \\ 
EL & Greece & 1.185 & 1.112 & 1.087 & 1.126 & 1.111 & 1.097 & 1.077 \\ 
HU & Hungary & 0.731 & 0.688 & 0.617 & 0.639 & 0.640 & 0.593 & 0.561 \\ 
IE & Ireland & 0.805 & 0.812 & 0.784 & 0.797 & 0.818 & 0.818 & 0.796 \\ 
IT & Italy & 0.964 & 0.993 & 0.944 & 0.904 & 0.942 & 1.026 & 0.963 \\ 
LV & Latvia & 0.507 & 0.447 & 0.680 & 0.569 & 0.654 & 1.002 & 0.683 \\ 
LT & Lithuania & 0.506 & 0.424 & 0.511 & 0.557 & 0.579 & 0.528 & 0.564 \\ 
LU & Luxembourg & 0.647 & 0.635 & 0.623 & 0.616 & 0.696 & 0.832 & 0.828 \\ 
MT & Malta & 0.711 & 0.653 & 0.612 & 0.478 & 0.551 & 0.502 & 0.524 \\ 
NL & Netherlands & 1.073 & 1.142 & 1.128 & 1.123 & 1.057 & 1.027 & 0.991 \\ 
PL & Poland & 0.731 & 0.661 & 0.642 & 0.609 & 0.619 & 0.570 & 0.612 \\ 
PT & Portugal & 0.699 & 0.663 & 0.623 & 0.684 & 0.644 & 0.698 & 0.735 \\ 
RO & Romania & 0.495 & 0.443 & 0.552 & 0.612 & 0.530 & 0.536 & 0.683 \\ 
SK & Slovakia & 0.826 & 0.705 & 0.778 & 0.655 & 0.691 & 0.599 & 0.669 \\ 
SI & Slovenia & 0.864 & 0.888 & 0.842 & 0.817 & 0.780 & 0.950 & 0.748 \\ 
ES & Spain & 0.674 & 0.697 & 0.734 & 0.722 & 0.746 & 0.822 & 0.914 \\ 
SE & Sweden & 1.041 & 1.068 & 0.997 & 0.978 & 1.036 & 1.005 & 0.977 \\ 
UK & United Kingdom & 0.849 & 0.859 & 0.845 & 0.858 & 0.835 & 0.844 & 0.818 \\ 
\hline
\end{tabular}
\caption{Efficiencies of general research of European Union countries in 2008--2014.}
\label{tab:effGeneral}
\end{center}
\end{table}

\begin{table}
\begin{center}
\begin{tabular}{llrrrrrrr}
\hline
Code & Country & 2008 & 2009 & 2010 & 2011 & 2012 & 2013 & 2014 \\
\hline
AT & Austria & 0.838 & 0.874 & 0.747 & 0.684 & 0.678 & 0.696 & 0.642 \\ 
BE & Belgium & 0.733 & 0.755 & 0.758 & 0.725 & 0.793 & 0.832 & 0.757 \\ 
BG & Bulgaria & 0.763 & 0.636 & 0.575 & 0.518 & 0.509 & 0.378 & 0.426 \\ 
HR & Croatia & 0.651 & 0.758 & 0.790 & 0.894 & 0.930 & 0.772 & 0.882 \\ 
CY & Cyprus & 0.996 & 1.124 & 1.141 & 1.129 & 1.258 & 1.259 & 1.310 \\ 
CZ & Czech Republic & 0.524 & 0.585 & 0.605 & 0.626 & 0.567 & 0.539 & 0.518 \\ 
DK & Denmark & 1.067 & 1.027 & 1.173 & 1.153 & 1.103 & 0.989 & 1.034 \\ 
EE & Estonia & 0.785 & 0.696 & 0.765 & 0.674 & 0.581 & 0.594 & 0.605 \\ 
FI & Finland & 0.759 & 0.773 & 0.764 & 0.715 & 0.727 & 0.841 & 0.799 \\ 
FR & France & 0.649 & 0.676 & 0.623 & 0.593 & 0.567 & 0.628 & 0.520 \\ 
DE & Germany & 0.594 & 0.625 & 0.572 & 0.531 & 0.480 & 0.597 & 0.532 \\ 
EL & Greece & 1.185 & 1.112 & 1.087 & 1.126 & 1.109 & 1.097 & 1.077 \\ 
HU & Hungary & 0.654 & 0.643 & 0.568 & 0.564 & 0.533 & 0.494 & 0.489 \\ 
IE & Ireland & 0.711 & 0.759 & 0.732 & 0.753 & 0.762 & 0.778 & 0.733 \\ 
IT & Italy & 0.954 & 0.993 & 0.914 & 0.899 & 0.915 & 1.026 & 0.963 \\ 
LV & Latvia & 0.348 & 0.317 & 0.546 & 0.427 & 0.354 & 0.348 & 0.421 \\ 
LT & Lithuania & 0.503 & 0.424 & 0.482 & 0.510 & 0.440 & 0.333 & 0.371 \\ 
LU & Luxembourg & 0.345 & 0.509 & 0.471 & 0.526 & 0.619 & 0.832 & 0.828 \\ 
MT & Malta & 0.645 & 0.450 & 0.538 & 0.478 & 0.322 & 0.376 & 0.402 \\ 
NL & Netherlands & 0.921 & 0.953 & 0.957 & 0.943 & 0.788 & 0.809 & 0.719 \\ 
PL & Poland & 0.720 & 0.653 & 0.566 & 0.551 & 0.472 & 0.439 & 0.455 \\ 
PT & Portugal & 0.686 & 0.661 & 0.622 & 0.684 & 0.635 & 0.697 & 0.733 \\ 
RO & Romania & 0.494 & 0.441 & 0.548 & 0.592 & 0.481 & 0.479 & 0.604 \\ 
SK & Slovakia & 0.816 & 0.705 & 0.750 & 0.611 & 0.650 & 0.564 & 0.669 \\ 
SI & Slovenia & 0.826 & 0.881 & 0.842 & 0.817 & 0.780 & 0.950 & 0.748 \\ 
ES & Spain & 0.642 & 0.683 & 0.726 & 0.711 & 0.702 & 0.820 & 0.914 \\ 
SE & Sweden & 0.921 & 0.999 & 0.848 & 0.826 & 0.761 & 0.766 & 0.755 \\ 
UK & United Kingdom & 0.760 & 0.797 & 0.783 & 0.771 & 0.653 & 0.670 & 0.599 \\ 
\hline
\end{tabular}
\caption{Efficiencies of basic research of European Union countries in 2008--2014.}
\label{tab:effBasic}
\end{center}
\end{table}

\begin{table}
\begin{center}
\begin{tabular}{llrrrrrrr}
\hline
Code & Country & 2008 & 2009 & 2010 & 2011 & 2012 & 2013 & 2014 \\
\hline
AT & Austria & 0.971 & 0.985 & 0.928 & 0.878 & 0.918 & 0.840 & 0.785 \\ 
BE & Belgium & 0.802 & 0.777 & 0.798 & 0.778 & 0.803 & 0.748 & 0.683 \\ 
BG & Bulgaria & 0.244 & 0.200 & 0.197 & 0.267 & 0.386 & 0.384 & 0.391 \\ 
HR & Croatia & 0.244 & 0.175 & 0.257 & 0.184 & 0.226 & 0.210 & 0.143 \\ 
CY & Cyprus & 0.534 & 0.664 & 0.358 & 0.282 & 0.157 & 0.446 & 0.411 \\ 
CZ & Czech Republic & 0.315 & 0.300 & 0.313 & 0.358 & 0.354 & 0.331 & 0.299 \\ 
DK & Denmark & 0.930 & 0.823 & 1.049 & 1.125 & 0.962 & 0.879 & 0.825 \\ 
EE & Estonia & 0.529 & 0.586 & 0.535 & 0.379 & 0.246 & 0.283 & 0.269 \\ 
FI & Finland & 0.818 & 0.819 & 0.935 & 0.873 & 1.046 & 1.105 & 1.133 \\ 
FR & France & 0.776 & 0.788 & 0.756 & 0.805 & 0.840 & 0.825 & 0.765 \\ 
DE & Germany & 1.108 & 1.133 & 1.106 & 1.041 & 0.995 & 0.954 & 0.886 \\ 
EL & Greece & 0.277 & 0.248 & 0.196 & 0.282 & 0.364 & 0.365 & 0.325 \\ 
HU & Hungary & 0.473 & 0.477 & 0.449 & 0.500 & 0.502 & 0.488 & 0.416 \\ 
IE & Ireland & 0.585 & 0.613 & 0.544 & 0.608 & 0.597 & 0.598 & 0.550 \\ 
IT & Italy & 0.900 & 0.852 & 0.898 & 0.873 & 0.889 & 0.809 & 0.783 \\ 
LV & Latvia & 0.477 & 0.411 & 0.504 & 0.459 & 0.593 & 1.002 & 0.544 \\ 
LT & Lithuania & 0.195 & 0.105 & 0.210 & 0.252 & 0.373 & 0.420 & 0.410 \\ 
LU & Luxembourg & 0.647 & 0.595 & 0.568 & 0.598 & 0.675 & 0.627 & 0.622 \\ 
MT & Malta & 0.480 & 0.637 & 0.351 & 0.033 & 0.456 & 0.337 & 0.319 \\ 
NL & Netherlands & 0.956 & 0.984 & 0.933 & 1.015 & 1.000 & 0.977 & 0.927 \\ 
PL & Poland & 0.342 & 0.394 & 0.416 & 0.400 & 0.495 & 0.457 & 0.456 \\ 
PT & Portugal & 0.222 & 0.150 & 0.143 & 0.187 & 0.211 & 0.229 & 0.228 \\ 
RO & Romania & 0.138 & 0.110 & 0.152 & 0.258 & 0.302 & 0.332 & 0.419 \\ 
SK & Slovakia & 0.371 & 0.290 & 0.434 & 0.400 & 0.353 & 0.315 & 0.241 \\ 
SI & Slovenia & 0.716 & 0.614 & 0.538 & 0.559 & 0.560 & 0.570 & 0.508 \\ 
ES & Spain & 0.402 & 0.416 & 0.412 & 0.426 & 0.493 & 0.502 & 0.471 \\ 
SE & Sweden & 1.020 & 0.996 & 0.997 & 0.973 & 1.017 & 1.000 & 0.963 \\ 
UK & United Kingdom & 0.616 & 0.652 & 0.648 & 0.693 & 0.749 & 0.744 & 0.672 \\ 
\hline
\end{tabular}
\caption{Efficiencies of applied research of European Union countries in 2008--2014.}
\label{tab:effApplied}
\end{center}
\end{table}

\subsection{Comparison of the Resulting Efficiencies with Other Studies}
\label{sec:empComp}

Comparison of our results with other studies is difficult for the following reason. One of the assumptions of DEA is the homogeneity of DMU set. However, most studies evaulating R\&D efficiency use rather heterogeneous set of countries \citep{Wang2007a, Sharma2008, Cincera2009, Chen2011}. This makes their efficiencies unreliable because DEA ranks DMUs relatively to other DMUs. \cite{Lee2005} overcome this issue by clustering countries to homogeneous sets of DMUs. If we compare the results despite homogeneity violation, the major differences often occur in countries like Croatia, Czech Republic, Hungary, Luxembourg and Romania \citep{Aristovnik2012, Cincera2009, Chen2011, Ekinci2017}.

\cite{Cincera2009} classify Croatia, Lithuania, Luxembourg, Malta and Romania as efficient countries. According to our study, they are inefficient. The possible reason for this is inclusion of countries from South America, North America, Europe and Asia in their study.

The study of \cite{Chen2011} conducted for the years 1998--2005 finds that Hungary is efficient while in our study it is inefficient. Again, this could be because they use heterogeneous set of 16 European countries, 4 Asian countries and 4 countries from South and North America.

\cite{Ekinci2017} use the homogeneous set of all 28 members of the European Union. However, they do not use prices adjusted to purchasing power parities which could also create major differences between the studies. They find Austria, Croatia, Luxembourg, Poland and Romania to be efficient while in our study they are inefficient with efficiency lower than 0.8. On the contrary, they do not find Denmark and Greece efficient unlike our study. Our preliminary results for the year 2014 not accounting for purchasing power parities are very similar to the results of \cite{Ekinci2017}, specially in the case of efficiency of Luxembourg and Romania \citep{Holy2017k}.

\subsection{Results of Panel Regression}
\label{sec:empPanel}

To asses our hypotheses we adopt panel regression with random time effects and logistic function to model R\&D efficiencies. All parameteres of panel regression are statistically significant with p-value almost zero for t-tests. Our analysis does not indicate any violation of panel regression assumptions. However, it does not show the direction of causality between R\&D efficiency and GDP. The use of random time effects seems to be appropriate as it captures fluctuations in overall GDP level.

In the analysis, we consider Luxembourg to be an outlier due to its high GDP per capita. The reasons why Luxembourg has one of the highest GDP per capita in the world lies in the fact that this country is an international centre of banking and financial services and only few people working there are actual citizens.

The results are reported in Table \ref{tab:panel}. We stress the importance of correct interpretation of the results as any change in the efficency of one country could also affect the efficiencies of other countries. We focus mainly on differences in steepness of the curve for basic and applied research which is more than double for applied research. This might be because the outputs (and their quality) of basic research are more problematic to identify than in the case of applied research.

\begin{figure}
\begin{center}
\includegraphics[width=118mm]{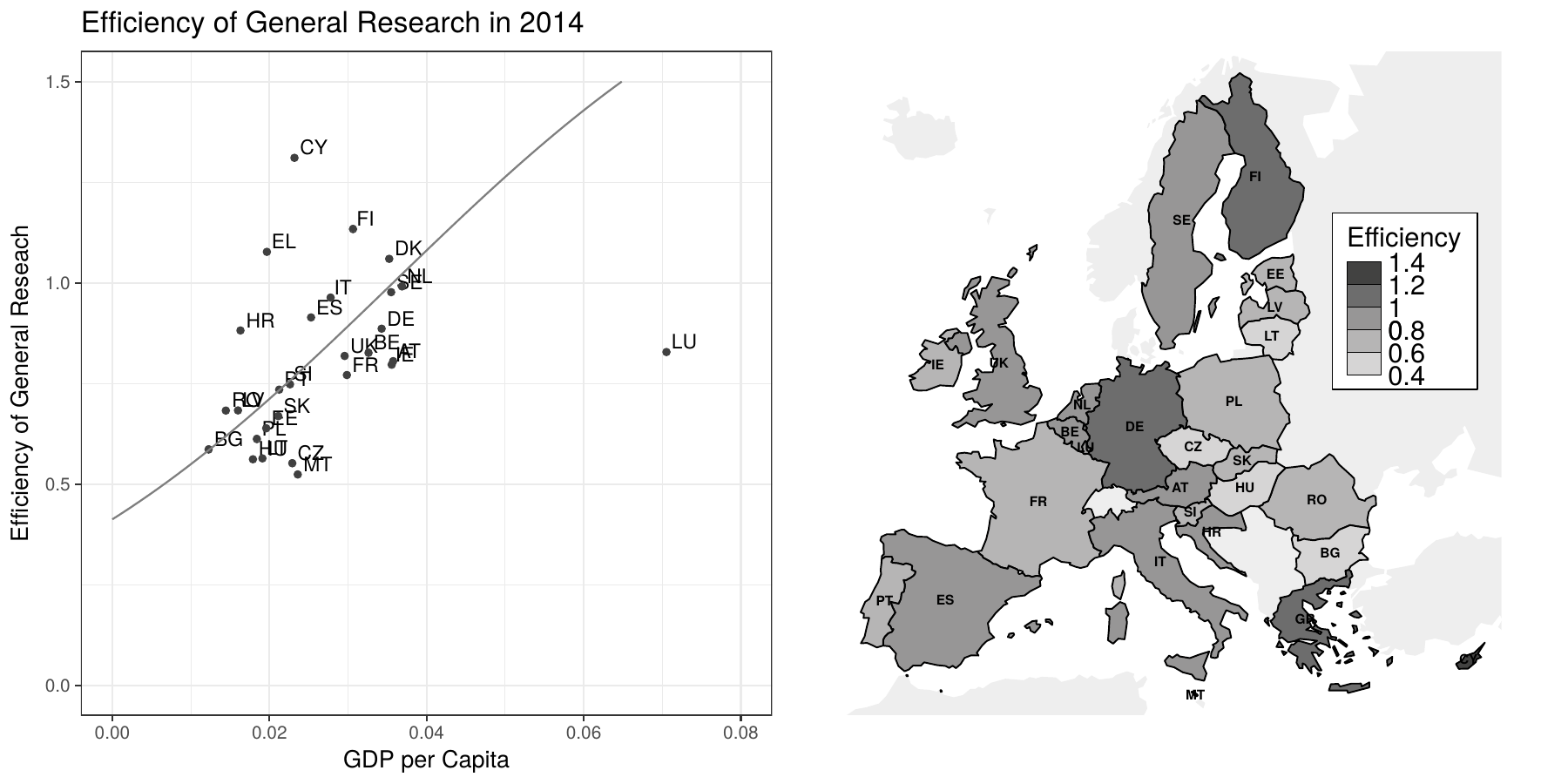} 
\caption{Relation of efficiency of general research and GDP per capita of European Union countries in 2014 (left) and the efficiencies of general research of European Union countries in 2014 (right).}
\label{fig:effGeneral}
\end{center}
\end{figure}

\begin{figure}
\begin{center}
\includegraphics[width=118mm]{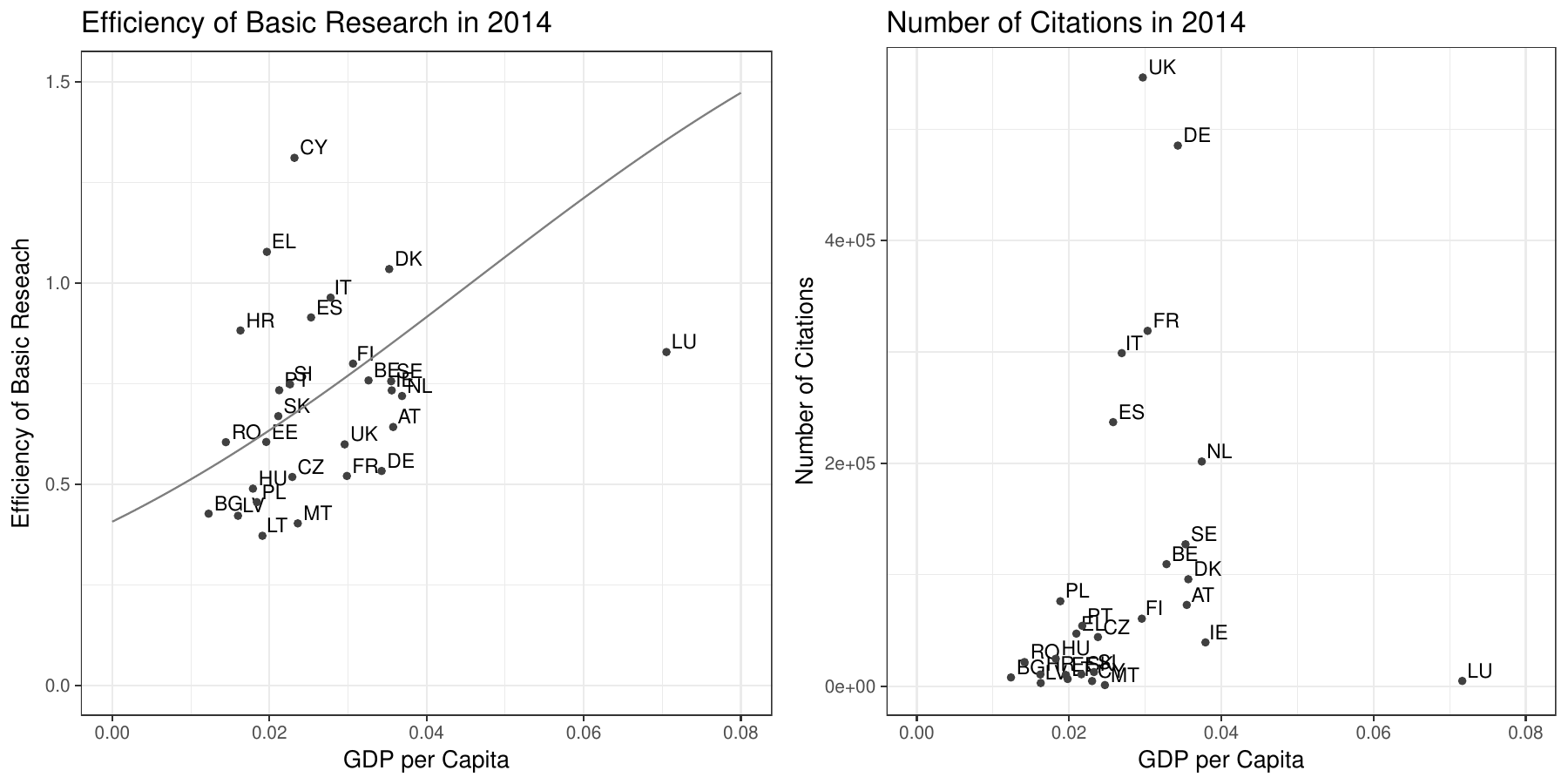} 
\caption{Relation of efficiency of basic research and GDP per capita of European Union countries in 2014 (left) and relation of the number of citations and GDP per capita of European Union countries in 2014 (right).}
\label{fig:effBasic}
\end{center}
\end{figure}

\begin{figure}
\begin{center}
\includegraphics[width=118mm]{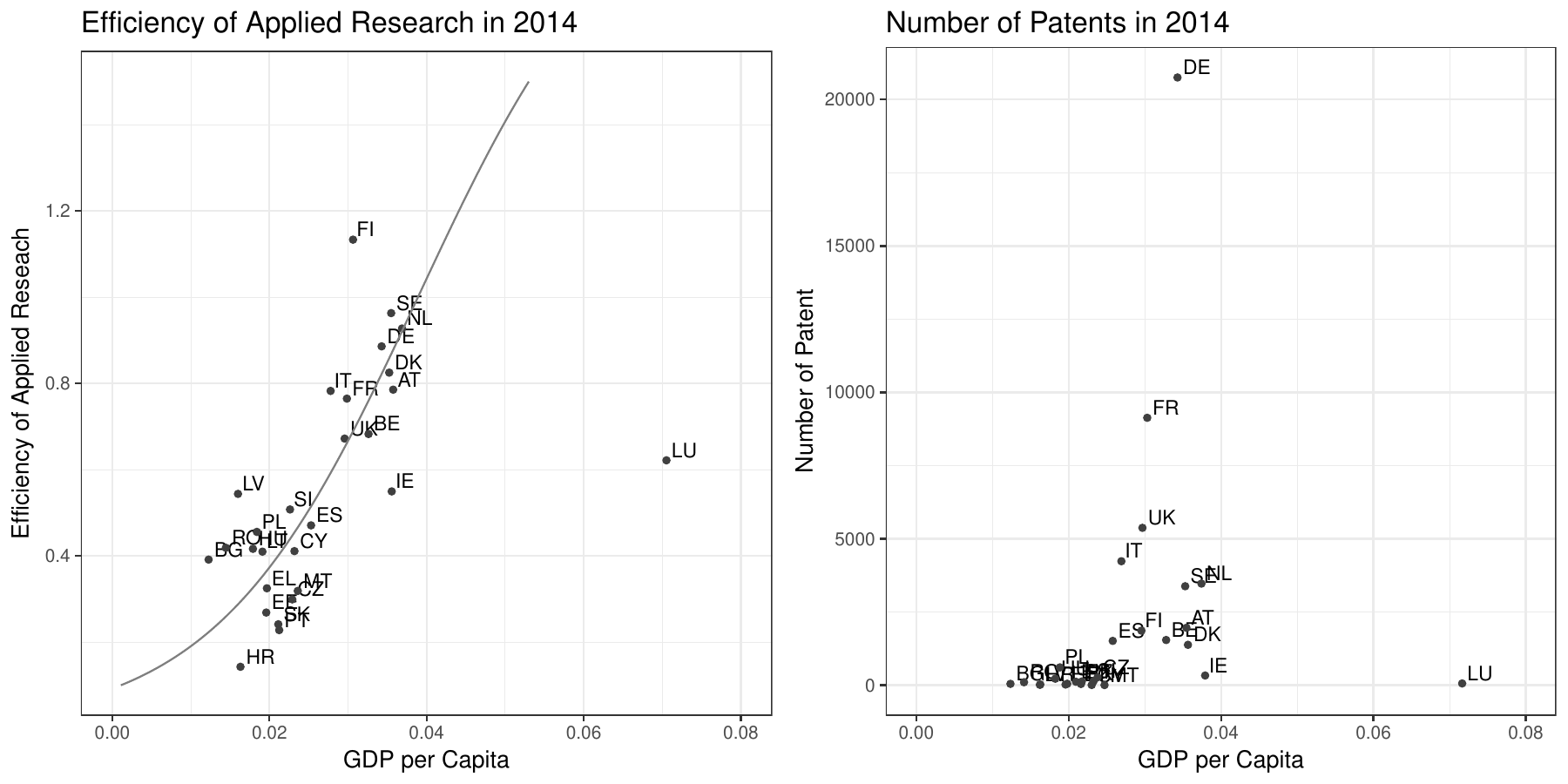} 
\caption{Relation of efficiency of applied research and GDP per capita of European Union countries in 2014 (left) and relation of the number of patents and GDP per capita of European Union countries in 2014 (right).}
\label{fig:effApplied}
\end{center}
\end{figure}

Figure \ref{fig:effGeneral} shows the relation of general research efficiency to GDP per capita with fitted logistic curve. The fitted logistic curve can also be used as another type of classification based on average efficiency for a given GDP per capita. For example the Czech Republic and Spain have similar GDP per capita and are both inefficient. The Czech Republic has efficiency 0.552 making it below average for its GDP per capita while Spain has efficiency 0.914 making it above average for its GDP per capita. Figure \ref{fig:effBasic} shows the relation of efficiency of basic research and the number of citations to GDP per capita. Interestingly, the number of citations does not exhibit any simple relation to GDP per capita. Countries with lower GDP per capita tend to have lower number of citations. However, countries with higher GDP are more volatile as some countries have higher number of citations and some lower. Similar behaviour can be seen in Figure \ref{fig:effApplied} for efficiency of applied research and the number of patents.

\begin{table}
\begin{center}
\begin{tabular}{lrrrrrrrr}
\hline
Research & $\hat{\alpha}$ & $\text{se}_{\hat{\alpha}}$ & $\text{t}_{\hat{\alpha}}$ & $\hat{\beta}$ & $\text{se}_{\hat{\beta}}$ & $\text{t}_{\hat{\beta}}$ & $\hat{\sigma}$ & $\hat{\omega}$ \\
\hline
General  & 1.348 & 0.083 & 16.226 & 37.734 & 3.086 & 12.227 & 0.042 & 0.316 \\
Basic    & 1.366 & 0.108 & 12.624 & 29.897 & 4.007 &  7.460 & 0.003 & 0.410 \\
Applied  & 3.029 & 0.146 & 20.761 & 77.866 & 5.281 & 14.744 & 0.111 & 0.538 \\
\hline
\end{tabular}
\caption{Results of panel regression models. The estimated intercept is denoted as $\hat{\alpha}$, its standard error as $\text{se}_{\hat{\alpha}}$ and its t-statistic as $\text{t}_{\hat{\alpha}}$. The estimated coefficient of GDP is denoted as $\hat{\beta}$, its standard error as $\text{se}_{\hat{\beta}}$ and its t-statistic as $\text{t}_{\hat{\beta}}$. Standard deviations are denoted as $\hat{\sigma}$ for time effects and $\hat{\omega}$ for idiosyncratic error terms.}
\label{tab:panel}
\end{center}
\end{table}

\subsection{Implications of the Results}
\label{sec:empImpl}

The results have direct implications on our hypotheses. The analysis shows that there exists statistically significant positive relation between the R\&D efficiency and the GDP per capita supporting Hypothesis 1. This is illustrated in Figure \ref{fig:effGeneral}.

Positive relation between specialized efficiencies is also statistically significant for both basic and applied research. In the case of applied research the dependency of R\&D efficiency on GDP per capita is much more stronger than in the case of basic research supporting Hypothesis 2. This is illustrated in Figure \ref{fig:effBasic} and Figure \ref{fig:effApplied}. Distribution of efficiencies of basic and applied research in Figure \ref{fig:hist} also supports  Hypothesis 2.

\section{Conclusion}
\label{sec:con}

We rank European Union countries based on R\&D research using the Chebyshev distance DEA methodology. We use the number of citations as output of basic research, the number of patents as output of applied research and R\&D expenditure with manpower as inputs. The main contributions of our paper are the following:
\begin{enumerate}
\item Unlike many other studies on similar topics, we take into account in our analysis the assumption of homogeneity between countries, purchasing power parity and time delay between inputs and outputs. 
\item We expose the positive relation between the R\&D efficiency and the level of economic advancement of a country. This is the expected result suggesting an appropriate strategic usage of variables.
\item We study distinctions between basic and applied research. We show that in the case of efficiency of applied research there are more significant differences between countries than in the case of basic research.
\end{enumerate}

We see the main application of our results in policy making. In the case of applied research, our results do not suggest omitting the number of patents as an indicator for outputs of applied research. In the case of basic research, efficiency of the number of citations is not as connected to the level of economic advancement of a country as some advanced countries demonstrate a rather low number of citations. This might be because the outputs of basic research are harder to capture.

\section*{Acknowledgements}
\label{sec:acknow}

We would like to thank Milan Hladík for his help with the Chebyshev distance DEA, Jakub Fischer for his comments and Alena Holá for proofreading.

\section*{Funding}
\label{sec:fund}

This work was supported by the Czech Science Foundation under the project DYME - Dynamic Models in Economics, No. P402/12/G097.


\end{document}